\documentclass[11pt]{amsart}

\newtheorem{Theorem}{Theorem}
\newtheorem{Corollary}{Corollary}
\newtheorem{Definition}{Definition}

\newtheorem{Lemma}{Lemma}

\newcommand{\Id}{{\mathrm{I}}}

\begin{document}

\title[Dynamical Lower Bounds for 1D Dirac Operators] {Dynamical Lower Bounds for 1D Dirac Operators}
\author{Roberto A. Prado}
\thanks{E-mail addresses: rap@dm.ufscar.br (RAP), oliveira@dm.ufscar.br
(CRdeO)}
\thanks{Fax: +55 16 33518218}
\thanks{RAP was supported by FAPESP (Brazil)}
\address{Departamento de Matem\'{a}tica -- UFSCar, S\~{a}o Carlos, SP,
13560-970 Brazil\\}
\email{rap@dm.ufscar.br}
\author{C\'{e}sar R. de Oliveira}
\thanks{CRdeO was partially supported by CNPq (Brazil)}
\address{Departamento de Matem\'{a}tica -- UFSCar, S\~{a}o Carlos, SP,
13560-970 Brazil}
\email{oliveira@dm.ufscar.br}
\subjclass{81Q10}

\begin{abstract}  Quantum dynamical lower bounds for
continuous and discrete one-dimensional
Dirac operators are established in terms of transfer matrices. Then such
results are applied to
various models, including the
Bernoulli-Dirac one and, in contrast to the discrete case, critical
energies are also found  for
 the continuous Dirac case with positive mass.
\end{abstract}
\maketitle

\section{Introduction} We consider discrete, resp.\ continuous, Dirac
operators
\begin{equation} \label{DiracOperator}
\textbf{D}(m,c):=\textbf{D}_{0}(m,c)+ V \Id_2 =
\left(\begin{array}{cc} mc^2 & cD^* \\ cD & -mc^2
\end{array} \right) + V \Id_2,
\end{equation} with Dirichlet boundary conditions, acting on
$\ell^2(\mathbb{N},\mathbb{C}^2)$, resp.\ $L^2([0,\infty),\mathbb{C}^2)$,
where $c>0$ represents the
speed of  light,
$m\geq 0$ the mass of a particle, $\Id_2$ is the $2\times2$ identity
matrix and $V$ is a bounded real
potential. In the discrete case $D$ is the finite difference operator
defined by
$(D\varphi)(n)=\varphi(n+1)-\varphi(n)$, with adjoint
$(D^{*}\varphi)(n)=\varphi(n-1)-\varphi(n)$, and in the continuous case
$D=D^{*}=-i\frac{d}{dx}$.

Model~(\ref{DiracOperator}) in the continuous case is well known in
relativistic quantum
mechanics \cite{BD,T}, and  the discrete version was introduced and
studied in \cite{dOP1,dOP2}.

The goal of this paper is to establish lower bounds on the dynamics
associated to $\textbf{D}(m,c)$
through  the behaviour of the corresponding transfer matrices. To this end
we will consider the time
averaged $q$-th moments $A_{\psi}$ of the position operator
\[
\Big[X\Big(\begin{array}{c} \varphi_+ \\ \varphi_- \end{array}\Big)\Big](x)=
\Big(\begin{array}{c} x\ \varphi_+(x) \\ x\ \varphi_-(x) \end{array}\Big)
\] acting in
$\ell^2(\mathbb{N},\mathbb{C}^2)$, resp.\ $L^2([0,\infty),\mathbb{C}^2)$,
defined by ($T>0$)
\begin{equation} \label{dynamic moments} A_{\psi}(m,T,q):= \frac{2}{T}
\int_0^{\infty}{e^{-2t/T}}
\left\| \left|X\right|^{q/2}  e^{-it\textbf{D}(m,c)}\psi \right\|^2\ dt ,
\end{equation} with initial state $\psi=\delta_{1}^+$ in
$\ell^2(\mathbb{N},\mathbb{C}^2)$, resp.\
$\psi=f$ in $L^2([0,\infty),\mathbb{C}^2)$, where $\delta_1^+$ is an
element of the canonical basis of
$\ell^2(\mathbb{N},\mathbb{C}^2)$ and
$f$ is an element of $L^2([0,\infty),\mathbb{C}^2)$ with compact support
which satisfies a suitable
technical condition.

To investigate the polynomial behaviour in time $T$ of $A_{\psi}(m,T,q)$,
one usually considers the lower growth
exponents
\begin{equation} \label{Transp.Exponents}
\beta_{\psi}^-(m,q):=\liminf_{T \to \infty}
\frac{\log A_{\psi}(m,T,q)}{\log T}\ .
\end{equation}

In the Schr\"odinger setting, dynamical lower bounds was found for random
polymer models \cite{JSBS} and
for random palindrome models
\cite{CadO}, due to existence of critical energies \cite{JSBS}. For discrete
Schr\"odinger operators in
$\ell^2(\mathbb{N})$ and $\ell^2(\mathbb{Z})$, in
\cite{DST} a  general method was developed which allows one to derive
dynamical lower bounds from upper
bounds on the growth of norms of transfer matrices. Damanik, Lenz and
Stolz \cite{DLS} have presented an
extension of this method to continuous Schr\"odinger operators in
$L^2([0,\infty))$ and $L^2(\mathbb{R})$, with application to the
continuous Bernoulli-Anderson model.

In this paper we adapt the above  mentioned methods to the Dirac model
(\ref{DiracOperator}) for both
 discrete and continuous cases. One important consequence of
Theorem~\ref{DynamicTeor} ahead is
the following: suppose that there is an energy $E_0\in\mathbb{R}$ such
that the transfer matrices
$\Phi_{m}(E_0,x,y)$ (defined in Section~\ref{DynamicSection}) satisfies
$\left\|\Phi_{m}(E_0,x,y)\right\|\leq CN^{\alpha}$ for all $N$ large
enough, $\alpha\geq 0$,
$C>0$ and
$0\leq x,y\leq N$, then it follows that
\[A_{\psi}(m,T,q)\geq \tilde{C} T^{\frac{q-1-4\alpha}{1+\alpha}} \ ,\] for
$\psi$ as in (\ref{dynamic
moments}) and $\tilde C>0$. We then apply such result to the continuous
Bernoulli-Dirac model,  the
discrete Dirac model with zero mass ($m=0$) and any two-valued potential,
the Thue-Morse Dirac
model and discrete Dirac model with Sturmian potentials.

There are some reasons justifying the adaptation of known results in the
Schr\"odinger setting to the
Dirac one. First of all, although expected, it is not immediately clear
(nor trivial) which and how such
adaptations work. Second, although we have found the abstract results have
similar statements, in
applications usually different conditions on the potentials appear in case
of Dirac operators (see, e.g., Theorem~\ref{thmabqany}). Third,
and this was our main motivation for considering dynamical lower bounds
for  model (\ref{DiracOperator}),
is that for the continuous Bernoulli-Dirac model  it is possible
to construct examples (see Subsection~\ref{subsecCBDM}) which have
critical energies for $m=0$ and also for
$m>0$, in contrast with the discrete case which have critical energies
only for $m=0$ \cite{dOP1,dOP2}.
Fourth, with respect to transfer matrices, the discrete Dirac operator has
some kind of ``built-in
dimerization'' \cite{dOP2} (implying transport) which motivates the study
of the corresponding
continuous case. Finally, we have found that the upper and lower
components of
some initial conditions in the
Dirac setting produce interferences so that the technique in the
Schr\"odinger case does not  apply
(so leaving an interesting open problem); see the remark at the end of
Subsection~\ref{subsecCBDM}.

We anticipate that the presence of critical
energies in continuum Bernoulli-Dirac models produces dynamical lower
bounds in the sense that almost
surely
\[\beta_{f}^{-}(m,q)\geq q-\frac{1}{2},\] for all $q>0$, for any mass
$m\geq 0$ and suitable initial
conditions $f$.

Another method to obtain dynamical lower bounds from upper bounds on
transfer matrices was lately
developed in~\cite{GKT}, with application to Schr\"odinger
operators with random decaying potentials and sparse potentials. Their
method is suitable for models
that admit upper bounds on transfer matrix norms for large sets of
energies (i.e., sets with positive
Lebesgue measure), while with the method used here (based on
\cite{DLS,DST}) it is possible to get
dynamical bounds for models with large or small (e.g., finite) sets of
such energies.
An approach for quasi-ballistic dynamics for discrete Schr\"odinger as
well Dirac operators with
potentials along some dynamical systems have recently been obtained
in~\cite{deOPqb}.

This paper is organized as follows: In Section~\ref{DynamicSection} the
result about dynamical lower
bounds (Theorem~\ref{DynamicTeor}) for the Dirac model
(\ref{DiracOperator}) is presented, whose proof
appears in Section~\ref{ProofsSection}. In Section~\ref{ApplicSection}
applications of
Theorem~\ref{DynamicTeor} are discussed, including the continuous
Bernoulli-Dirac model.

\

\section{Dynamical Bounds}
\label{DynamicSection} In this section we will present  results about
dynamical lower bounds for the
operators $\textbf{D}(m,c)$ defined by (\ref{DiracOperator}) in both the
discrete and continuous cases.

For a given operator $\textbf{D}(m,c)$ on
$\ell^2(\mathbb{N},\mathbb{C}^2)$, resp.\ $L^2([0,\infty),\mathbb{C}^2)$,
the transfer matrices
$\Phi_m(E,x,y)$ between sites $y$ and
$x$ are defined  as
$$\Phi_m(E,x,y)=\left(\begin{array}{cc} u_{+}^{N}(x+1) & u_{+}^{D}(x+1) \\
  u_{-}^{N}(x) & u_{-}^{D}(x) \end{array} \right),\ \mbox{resp.}\
  \left(\begin{array}{cc} u_{+}^{N}(x) & u_{+}^{D}(x) \\
  u_{-}^{N}(x) & u_{-}^{D}(x) \end{array} \right), $$ where
$u^{N}=\left(\begin{array}{c} u_{+}^{N} \\ u_{-}^{N} \end{array}\right)$ and
$u^{D}=\left(\begin{array}{c} u_{+}^{D} \\ u_{-}^{D} \end{array}\right)$
denote the solutions of
equation $\textbf{D}(m,c)u=Eu$, $E\in\mathbb{R}$, satisfying
$$\left(\begin{array}{c} u_{+}^{N}(y+1) \\ u_{-}^{N}(y) \end{array}\right)=
  \left(\begin{array}{c} 1 \\ 0 \end{array}\right),\
  \left(\begin{array}{c} u_{+}^{D}(y+1) \\ u_{-}^{D}(y) \end{array}\right)=
  \left(\begin{array}{c} 0 \\ 1 \end{array}\right),$$ resp.\
$$u^{N}(y)=\left(\begin{array}{c} 1 \\ 0 \end{array}\right),\
  u^{D}(y)=\left(\begin{array}{c} 0 \\ 1 \end{array}\right).$$ It follows
from the definitions that if
$u=\left(\begin{array}{c} u_{+} \\ u_{-} \end{array}\right)$ is a solution
of the eigenvalue equation
$\textbf{D}(m,c)u=Eu$, then
$$\left(\begin{array}{c} u_{+}(x+1) \\ u_{-}(x) \end{array}\right)=
  \Phi_m(E,x,y)
  \left(\begin{array}{c} u_{+}(y+1) \\ u_{-}(y) \end{array}\right),$$
resp.\ $$\left(\begin{array}{c} u_{+}(x) \\ u_{-}(x) \end{array}\right)=
  \Phi_m(E,x,y)
  \left(\begin{array}{c} u_{+}(y) \\ u_{-}(y) \end{array}\right).$$ Note
that in the discrete case, the
matrix  $\Phi_m(E,x,y)$, $x>y\geq 0$, can be written as
\[\Phi_{m}(E,x,y)=T_m(E,V(x))\cdots T_m(E,V(y+1)),\] with
\[ T_{m}(E,V(k))=\left(
\begin{array}{cc} 1+\displaystyle\frac{m^2c^4-(E-V(k))^2}{c^2} &
\displaystyle\frac{mc^2 +E-V(k)}{c}\\ \\
\displaystyle\frac{mc^2-E+V(k)}{c} & 1 \\
\end{array} \right).\]

We denote by $\delta_n^{\pm}$ the elements of the canonical  position
basis of
$\ell^2(\mathbb{N},\mathbb{C}^2)$, for which all entries are
$\left(\begin{array}{c} 0 \\ 0 \\ \end{array}\right)$ except  the $n$th
one, which is given by
$\left(\begin{array}{c} 1 \\ 0
\\ \end{array}\right)$ and $\left(\begin{array}{c} 0 \\ 1 \\
\end{array}\right)$ for the superscript indices~$+$ and~$-$, respectively.

In the continuous case, consider the measurable locally bounded
vector-valued functions
$w_E,v_E$ defined by
\[w_E(x)=u_+^N(0)
\left(\begin{array}{c} -u_+^D(x) \\ \\ u_-^D(x) \\ \end{array}\right)
+u_+^D(0)
\left(\begin{array}{c} u_+^N(x) \\ \\ -u_-^N(x) \\ \end{array}\right)\] and
\[v_E(x)=u_-^N(0)
\left(\begin{array}{c} -u_+^D(x) \\ \\ u_-^D(x) \\ \end{array}\right)
+u_-^D(0)
\left(\begin{array}{c} u_+^N(x) \\ \\ -u_-^N(x) \\ \end{array}\right).\]

For $g=\left(\begin{array}{c} g_+ \\ g_- \\ \end{array}\right),$ with
$g_+,g_-$ measurable and locally
bounded functions, and
    $ f=\left(\begin{array}{c} f_+ \\ f_- \\ \end{array}\right)
     \in L^2([0,\infty),\mathbb{C}^2)$ of compact support,  define
\[[ g, f ] := \int^{\infty}_{0}
\left(\overline{g_+(t)}\, f_+(t)+ \overline{g_-(t)}\, f_-(t)\right) dt.\]
Note that in case all involved functions are square integrable
$[\cdot,\cdot]$ coincides with their inner
product.

For fixed parameters $m$ and $c$, let $\mathcal{H}_E$ be the set of the
vectors
$f=\left(\begin{array}{c} f_+ \\ f_- \\ \end{array}\right)
     \in L^2([0,\infty),\mathbb{C}^2)$ with compact support, which satisfies
one of the following conditions:
\begin{itemize}
\item[(i)] $f_+\neq 0,\ f_- =0$ and
$[\overline{u},f]=\int^{\infty}_{0}u_+(t)f_+(t)dt \neq 0$
           for some solution
           $u=\left(\begin{array}{c} u_+ \\ u_- \\ \end{array}\right)$ of
           $\textbf{D}(m,c)u=Eu$;

\item[(ii)] $f_+ =0,\ f_-\neq 0$ and
$[\overline{u},f]=\int^{\infty}_{0}u_-(t)f_-(t)dt \neq 0$
           for some solution
           $u=\left(\begin{array}{c} u_+ \\ u_- \\ \end{array}\right)$ of
           $\textbf{D}(m,c)u=Eu$;

\item[(iii)] $f_+ \neq 0,\ f_-\neq 0$ and $[\overline{w_E}, f ] \neq 0$
             or $[\overline{v_E}, f ] \neq 0$ (or both).
\end{itemize}

\

For $\alpha,m \geq 0,\ C>0$ and $\ N>1$  define the set
\[P_m(\alpha,C,N)=
\bigg\{E\in\mathbb{R}: \left\|\Phi_{m}(E,x,y)\right\|\leq CN^{\alpha}
\ \mbox{for all}\ 0\leq x,y\leq N \bigg\}.\]

Now we are in position to state the main result about dynamical lower bounds.

\begin{Theorem} \label{DynamicTeor}
Let $\textbf{D}(m,c)$ be the operator defined by
(\ref{DiracOperator}). Suppose $E_0 \in \mathbb{R}$ is such that there
exist $C>0$ and $\alpha\geq 0$
with $E_0\in P_m(\alpha,C,N)$ for all sufficiently large $N$.

\noindent
$\bf (i)$ \rm{(discrete case)} Let $A(N)$ be a uniformly bounded sequence
of subset
of $P_m(\alpha,C,N)$
containing
$E_0$ and $\mu^{m}_{+}$ the spectral measure for
$\textbf{D}(m,c)$ associated to $\delta_1^+$.
Then, there exists $\tilde{C}>0$ such that for $T>0$ large enough
\[A_{\delta_1^+}(m,T,q)\geq \tilde{C}
\left(\left|B_2(T)\right|+\mu^{m}_{+}(B_1(T))\right)
T^{\frac{q-3\alpha}{1+\alpha}},\] where
$B_j(T), j=1,2,$ is the $j/T$ neighborhood of
$A(T^{\frac{1}{1+\alpha}})$.

\noindent
$\bf (ii)$ \rm{(continuous case)} Let $A(N)$ be a subset of
$P_m(\alpha,C,N)$ containing $E_0$ such that
$\mathrm{diam}(A(N))\rightarrow 0$ as $N\rightarrow\infty$. Then,
for every $f\in\mathcal{H}_{E_0}$ there exists $\tilde{C}>0$
such that for $T>0$ large enough
\[A_{f}(m,T,q)\geq \tilde{C}\left|B_1(T)\right|
T^{\frac{q-3\alpha}{1+\alpha}}.\]
\end{Theorem}

\

\noindent{\it Remarks.}
\noindent
$\bf 1.$ Theorem~\ref{DynamicTeor} can be adapted to the operator
$\textbf{D}(m,c)$ on $\ell^2(\mathbb{Z},\mathbb{C}^2)$ and
$L^2(\mathbb{R},\mathbb{C}^2)$, and always with similar statements.

\noindent
$\bf 2.$ The dynamical lower bounds obtained in Theorem~\ref{DynamicTeor}
are stable under suitable
power-decaying perturbations of the potential $V$ as in~\cite{DST},
because the power-law bounds of the
transfer matrices keep unchanged.

\

The proof of Theorem~\ref{DynamicTeor} will be given in
Section~\ref{ProofsSection}. As
in~\cite{DLS,DST},  Theorem~\ref{DynamicTeor} have the following immediate
consequences.

\begin{Corollary} \label{DynamicCor1}
Let $A$ be a nonempty bounded subset of
$P_m(\alpha,C,N)$ for some $C>0,\ \alpha\geq 0$ and for all $N$ large
enough, such that
$\mu^{m}_{+}(A)>0$. Then
\[\beta_{\delta_1^+}^-(m,q)\geq \frac{q-3\alpha}{1+\alpha}.\]
\end{Corollary}

\begin{proof} Take $A(N)=A$ for every $N$. Since
$\mu^{m}_{+}(B_1(T))\geq \mu^{m}_{+}(A)>0$, by
Theorem~\ref{DynamicTeor}(i) there
exists $\tilde{C}>0$ such that for $T>0$ large enough
\[A_{\delta_1^+}(m,T,q)\geq \tilde{C}\ T^{\frac{q-3\alpha}{1+\alpha}}.\]
Hence  the result follows.
\end{proof}
\

\begin{Corollary} \label{DynamicCor2}
Suppose there is an energy $E_0\in\mathbb{R}$ such that
$\left\|\Phi_{m}(E_0,x,y)\right\|\leq CN^{\alpha}$ for all
$N$ large enough and $0\leq x,y \leq N$. Then,
\[\beta_{\psi}^-(m,q)\geq \frac{q-1-4\alpha}{1+\alpha},\] for every
$\psi=f\in\mathcal{H}_{E_0}$ in the continuous case and
$\psi=\delta_1^+$ in the discrete case.
\end{Corollary}

\begin{proof} Take $A(N)=\{E_0\}$ for every $N$. Then
$B_1(T)=\left[E_0-\frac{1}{T},E_0+\frac{1}{T}\right]$ and by
Theorem~\ref{DynamicTeor} there exists
$\tilde{C}>0$ such that for $T$ large enough
\[A_{\psi}(m,T,q)\geq \frac{\tilde{C}}{T}\
T^{\frac{q-3\alpha}{1+\alpha}}=\tilde{C}\
T^{\frac{q-1-4\alpha}{1+\alpha}},\] for $\psi$ as in the hypothesis.
Hence the result follows.
\end{proof}

\

\section{Applications}
\

\label{ApplicSection} This section is devoted to applications of
Theorem~\ref{DynamicTeor} and its corollaries.

\

\subsection{The continuous Bernoulli-Dirac model} \label{subsecCBDM}
Let $g_0$ and $g_1$ be two
real-valued potentials with support in
$[0,1]$. Consider the family of Dirac operators in
$L^2([0,\infty),\mathbb{C}^2)$,
\begin{equation} \label{Dirac-Bernoulli}
\textbf{D}_{\omega}(m,c):=\textbf{D}_{0}(m,c)+ V_{\omega} \Id_2, \
\  \  \omega\in\Omega :=\left\{0,1\right\}^{\mathbb{N}},
\end{equation} with potential $V_{\omega}(x)=\sum_{n}g_{\omega_n}(x-n)$,
where $\omega_n\in \{0,1\}$ are
i.i.d.\ Bernoulli random variables with common probability measure $\mu$
satisfying $\mu(\{0\})=p$,
$\mu(\{1\})=1-p$, for some $0<p<1$, and product measure
$\textbf{P}=\prod_{n}\mu\left(\omega_n)\right)$ on $\Omega$.

Let $T_m^{(j)}(E)$ be the transfer matrix for
$\textbf{D}_{\omega}(m,c)$ with potential
$V_{j}(x)=\sum_{n}g_{j}(x-n)$, $j=0,1$, at energy $E$ from 0 to 1.

\

\begin{Definition} [\cite{JSBS}] \label{CriticDef}  $E_0\in\mathbb{R}$ is
a critical energy for
$\textbf{D}_{\omega}(m,c)$ if the matrices $T_m^{(j)}(E_0), \, j=0,1,$ are
elliptic (i.e.,
$|\mathrm{trace}\ T_m^{(j)}(E_0)|<2$) or equal to
$\pm \Id_2$, and commute.
\end{Definition}
\

If $E_0$ is a critical energy for $\textbf{D}_{\omega}(m,c)$, it follows
from Definition~\ref{CriticDef}
that there exists a real invertible matrix $Q$ such that
$$Q\ T_m^{(j)}(E_0)\ Q^{-1}=\left(\begin{array}{cc}
\cos(\eta_{j}) & -\sin(\eta_{j}) \\
\sin(\eta_{j}) & \cos(\eta_{j}) \\
\end{array}\right), \ \ \mbox{for}\ j=0,1.$$ Adapting the arguments used
in \cite{JSBS,DLS} for the
Bernoulli-Dirac model~(\ref{Dirac-Bernoulli}), we obtain the following
(details omitted).
\

\begin{Lemma} \label{Lim.Unif.Matrices} Assume that $\eta_0 -\eta_1$ is
not an integer multiple of $\pi$.
Let $\lambda>0$ be arbitrary. Then there are $b>0$ and $C<\infty$ such
that for every $N\in\mathbb{N},$
there exists a set
$\Omega_N(\lambda)\subset\Omega$ with
$\textbf{P}\left(\Omega_N(\lambda)\right)\leq Ce^{-bN^{\lambda}}$ and
\[\big\|\Phi_m^{\omega}(E,x,y)\big\| \leq C \] for all
$\omega\in\Omega\backslash\Omega_N(\lambda)$, \
$0\leq x,y \leq N$ and
$E\in [E_0-N^{-\lambda-1/2},E_0+N^{-\lambda-1/2}] .$
\end{Lemma}
\

We can now state our main result for  model~(\ref{Dirac-Bernoulli}).
\

\begin{Theorem} \label{BernoDynamicTeor} Assume that $\eta_0 -\eta_1$ is
not an integer multiple of
$\pi$. For every $f\in\mathcal{H}_{E_0}$ one has
\[\beta_{f}^-(m,q)\geq q-\frac{1}{2}\  ,\  \  \  \omega\
\textbf{P}-a.s.\ . \]
\end{Theorem}

\begin{proof} Due to Lemma~\ref{Lim.Unif.Matrices}, for each $\lambda>0$,
$\textbf{P}\left(\Omega_N(\lambda)\right)$ is summable over $N$. Thus, by
Lemma~\ref{Lim.Unif.Matrices}
and a Borel-Cantelli argument, there exists $0<C<\infty$ such that
$\|\Phi_m^{\omega}(E,x,y)\|\leq C$ \ for all $N$, $0\leq x,y\leq N$,
for almost every $\omega$ and
$E\in A(N):=[E_0-N^{-\lambda-1/2},E_0+N^{-\lambda-1/2}]$. Note that
$|B_1(T)|\geq |A(T)|=2 T^{-\lambda -1/2}$. Applying
Theorem~\ref{DynamicTeor}(ii) with $\alpha=0$, it follows that almost surely
$\beta_{f}^-(m,q)\geq q-\frac{1}{2}-\lambda$ for every
$f\in\mathcal{H}_{E_0}$. Taking
$\lambda=\frac{1}{n}\rightarrow 0$ and using a countable intersection of
full measure sets, we obtain the result.
\end{proof}

\

It is possible to show, by applying similar arguments of \cite{DLS,JSBS} for
model~(\ref{Dirac-Bernoulli}),
that if $E_0$ is a critical energy for $\textbf{D}_{\omega}(m,c)$, then
for every $f\in\mathcal{H}_{E_0}$ one has $\beta_{f}^-(m,q)\geq q-1$,
for every $\omega$.

Recently, we have established  (see~\cite{dOP2}) the same lower bounds
obtained above for the discrete
Bernoulli-Dirac model with zero mass $(m=0)$, due to existence of critical
energies. Now we will present
a continuous Bernoulli-Dirac model defined by~(\ref{Dirac-Bernoulli})
that have critical energies for both
$m=0$ and  $m>0$ (note that for the latter case critical energies are
absent in the discrete case). As a
consequence we will obtain lower bounds by Theorem~\ref{BernoDynamicTeor}.

In fact, consider the Bernoulli-Dirac model (\ref{Dirac-Bernoulli})
with $g_0=0$ and $g_1=\lambda
\chi_{[0,1]},\ \lambda>0$. By solving the equation
$\textbf{D}_0(m,c)u=Eu$ one finds the following
solutions for $E^2>m^2c^4$:
$u^{N}=\left(\begin{array}{c} u_{+}^{N} \\ u_{-}^{N} \end{array}\right)$
and
$u^{D}=\left(\begin{array}{c} u_{+}^{D} \\ u_{-}^{D} \end{array}\right)$,
with

\[u_{+}^{N}(x)=\cos (\xi_E\ x), \  \
u_{-}^{N}(x)=\displaystyle\frac{-i(mc^2-E)}{c \xi_E}
             \sin (\xi_E\ x),\]

\[u_{+}^{D}(x)=\displaystyle\frac{-i c \xi_E}{mc^2-E}
\sin (\xi_E\ x), \  \  u_{-}^{D}(x)=\cos (\xi_E\ x),\]
where \ $\xi_E=\displaystyle\frac{\sqrt{E^2-m^2c^4}}{c}$,
and they satisfy
\[u^{N}(0)=\left(\begin{array}{c} 1 \\ 0 \end{array}\right)
\  \  \mbox{and} \  \
u^{D}(0)=\left(\begin{array}{c} 0 \\ 1 \end{array}\right).\]

Thus, the transfer matrices are
\[ T_{m}^{(0)}(E)=\left( \begin{array}{cc}
\cos \xi_E & \displaystyle\frac{-ic\ \xi_E}{mc^2-E}\sin \xi_E  \\ \\
\displaystyle\frac{-i(mc^2-E)}{c\ \xi_E}\sin \xi_E  & \cos \xi_E  \\
\end{array} \right)\]
for $E^2>m^2c^4$ and
$T_{m}^{(1)}(E)=T_{m}^{(0)}(E-\lambda)$ for
$(E-\lambda)^2>m^2c^4$.

If $E=\pm\sqrt{m^2c^4+n^2\pi^2c^2}$ for $n\in\mathbb{N}^*$ and
$m\geq 0$, then $T_{m}^{(0)}(E)=\pm \Id_2$. Moreover, taking
\[
0<\lambda<\sqrt{m^2c^4+n^2\pi^2c^2}-mc^2\;\; \mathrm{or}\;\;
\lambda>\sqrt{m^2c^4+n^2\pi^2c^2}+mc^2
\] (this implies
$(E-\lambda)^2>m^2c^4$), it follows that
$|\mbox{trace}\ T_{m}^{(1)}(E) |<2$ (i.e., $T_{m}^{(1)}(E)$ is elliptic).
On the other hand, if
$E=\lambda\pm\sqrt{m^2c^4+n^2\pi^2c^2}$ for $n\in\mathbb{N}^*$,
$m\geq 0$ and $\lambda$ as above, we have $T_{m}^{(1)}(E)=\pm \Id_2$ and
$|\mbox{trace}\ T_{m}^{(0)}(E)
|<2$. Thus, for
such values of $\lambda$ we have the following set of critical energies:
\[\left\{\pm\sqrt{m^2c^4+n^2\pi^2c^2},\
         \lambda\pm\sqrt{m^2c^4+n^2\pi^2c^2} :
         n\in\mathbb{N}^*, m\geq 0 \right\}.\] For such energies  the
condition required in
Theorem~\ref{BernoDynamicTeor} holds, that is,
$\eta_0 -\eta_1\neq k\pi,\ k\in\mathbb{Z}.$

\

\begin{Corollary} \label{DynamicCor3} Let $\textbf{D}_{\omega}(m,c)$ be
defined
by~(\ref{Dirac-Bernoulli}) with $g_0=0$ and $g_1=\lambda\chi_{[0,1]}$,
$\lambda>0$. If
$\lambda<\sqrt{m^2c^4+n^2\pi^2c^2}-mc^2$ or
$\lambda>\sqrt{m^2c^4+n^2\pi^2c^2}+mc^2$, then
\[\beta_{f}^-(m,q)\geq q-\frac{1}{2}\  ,\  \  \  \omega\
\textbf{P}-a.s., \] for all  masses $m\geq 0$ and any
$f=\left(\begin{array}{c} f_+ \\ f_- \\ \end{array}\right)
   \in L^2([0,1],\mathbb{C}^2)$ satisfying one of the following conditions:

\noindent (i) $0\neq f_+ \in L^2([0,1])$ and $f_- =0$.

\noindent (ii) $f_+ =0$ and $0\neq f_- \in L^2([0,1])$.

\noindent (iii) $f_+ \neq 0,\ f_-\neq 0$ and
    \[[\overline{w_E}, f ] =\int_{0}^{1}\left[\left(\frac{-in\pi c}
{mc^2\mp\sqrt{m^2c^4+n^2\pi^2c^2}}\right)f_+(x)\sin(n\pi x)+
    f_-(x)\cos(n\pi x) \right]dx\neq 0\] or
    \[[\overline{v_E}, f ] =\int_{0}^{1}\left[f_+(x)\cos(n\pi x)-i
    \left(\frac{mc^2\mp\sqrt{m^2c^4+n^2\pi^2c^2}}{n\pi c}\right)
    f_-(x)\sin(n\pi x) \right]dx\neq 0 .\]
Note that in this case the above conditions on $f$ depends on $m$.
\end{Corollary}

\begin{proof} We consider two cases:
\begin{itemize}
\item[1.]$\omega_{0}=0$, that is, $V_{\omega}(x)=0$ on $[0,1]$.
\item[2.]$\omega_{0}=1$, that is, $V_{\omega}(x)=\lambda$ on $[0,1]$.
\end{itemize} If $\omega_{0}=0$, then applying
Theorem~\ref{BernoDynamicTeor} for the critical energies
\[E=\pm\sqrt{m^2c^4+n^2\pi^2c^2},\ n\in\mathbb{N}^*,\ m\geq 0,\] we obtain
\[\beta_{f}^-(m,q)\geq q-\frac{1}{2}\  ,\  \  \  \omega\
\textbf{P}-a.s., \] for all mass values $m\geq 0$ and for any
$f\in\mathcal{H}_{E}$ with $\mathrm{supp\,}f\subset
[0,1]$. Note that for such energies
\[u^N(x)=\left(\begin{array}{c} \cos(n\pi x) \\ \\
       -i\displaystyle\frac{(mc^2\mp\sqrt{m^2c^4+n^2\pi^2c^2})}{n\pi c}
       \sin(n\pi x) \\ \end{array}\right)\] and
\[u^D(x)=\left(\begin{array}{c} \displaystyle\frac{-in\pi c}
       {mc^2\mp\sqrt{m^2c^4+n^2\pi^2c^2}} \sin(n\pi x) \\ \\
        \cos(n\pi x) \\ \end{array}\right)\] are fundamental solutions of
$\textbf{D}_0(m,c)u=Eu$. By
definition we have the vectors
$w_E(x)=\left(\begin{array}{c} -u_+^D(x) \\ u_-^D(x) \\
              \end{array}\right)$ and
$v_E(x)=\left(\begin{array}{c} u_+^N(x) \\ -u_-^N(x) \\
              \end{array}\right)$.

For any $f_+ \in L^2([0,1]),\ f_+\neq 0$, there is at least one
$n\in\mathbb{N}$ such that
\[\int^{1}_{0}f_+(t)\cos(n\pi x)dt \neq 0 \ \ \mbox{or} \ \
\int^{1}_{0}f_+(t)\sin(n\pi x)dt \neq 0 \] (similarly for $0\neq f_- \in
L^2([0,1])$). This is valid
because
\[\{1\}\cup \{\cos(2k\pi x),\sin(2k\pi x) : k\in \mathbb{N}\}\]
                  form a basis of $L^2([0,1])$. Therefore, by using the
definition of the set
$\mathcal{H}_{E}$  the required result is obtained.

If $\omega_0=1$, then we conclude the result in the same way, but now
based on the critical energies
$E=\lambda\pm\sqrt{m^2c^4+n^2\pi^2c^2}$,
$n\in\mathbb{N}^*$ and $m\geq 0$.
\end{proof}

\

\noindent{\it Remark.} Note that  Corollary~\ref{DynamicCor3}{\it (iii)}
does not assure $\beta_{f}^-(m,q)\geq q-\frac{1}{2}$ for any
$f=\left(\begin{array}{c} f_+ \\ f_- \\ \end{array}\right)
   \in L^2([0,1],\mathbb{C}^2)$, due to some kind of quantum interference.
For instance, for any integer
$n,\tilde n$, by taking
\[f_+(x)=\frac{mc^2\mp\sqrt{m^2c^4+n^2\pi^2c^2}}{in\pi c}
\sin(\tilde{n}\pi x) \  \  \mbox{and} \  \ f_-(x)=-\cos(\tilde{n}\pi x),\]
one obtains
$[\overline{w_E}, f ] =0$ and $[\overline{v_E}, f ] =0$. In the
corresponding Schr\"odinger model~\cite{DLS} one has
$\beta_{f}^-(q)\geq q-\frac{1}{2}$ for any $f\in L^2([0,1]), f\neq 0$.

\

\subsection{The discrete massless Dirac model with two-valued
potentials} Consider the discrete
Dirac operator
$\textbf{D}(0,c)$ defined by~(\ref{DiracOperator}). The following result
holds.

\

\begin{Theorem}\label{thmabqany} Let $V:\mathbb{N}\rightarrow
\{a,b\}\subset\mathbb{R}$ be
a potential  for
$\textbf{D}(0,c)$.
\begin{itemize}
\item[(i)] If $|a-b|<2c$, then for every $q>0$,
$\beta_{\delta_1^+}^-(0,q)\geq q-1$.
\item[(ii)] If $|a-b|=2c$, then for every $q>0$,
$\beta_{\delta_1^+}^-(0,q)\geq \frac{q-5}{2}$.
\end{itemize}
\end{Theorem}

\begin{proof} We shall find upper bounds for the transfer matrices
$\Phi_0(E_0,x,y)$ for a  suitable energy $E_0$. Let $E_0=a$. Then
\[T_0(E_0,a)=\Id_2 \  \  \  \mbox{and} \  \  \
  T_0(E_0,b)=\left( \begin{array}{cc}
  \displaystyle 1-\frac{(a-b)^2}{c^2} &
  \displaystyle \frac{a-b}{c} \\ \\
  \displaystyle \frac{-a+b}{c} & 1 \\
  \end{array} \right).\] This implies that
$\Phi_0(E_0,x,y)=\left(T_0(E_0,b)\right)^{n_b}$, where $n_b$
is the number of times that $b$ occurs  in the product. If $|a-b|<2c$,
then $T_0(E_0,b)$ is elliptic
$(|\mbox{trace}\ T_0(E_0,b)|<2)$ and hence
\[\|\Phi_0(E_0,x,y)\|\leq C(E_0),\  \  \forall\ x,y\in\mathbb{N}.\] Thus,
by Corollary~\ref{DynamicCor2}
with $\alpha =0$, we obtain
\[\beta_{\delta_1^+}^-(0,q)\geq q-1 ,\  \  \forall\ q>0.\] On the other
hand, if $|a-b|=2c$, then
$T_0(E_0,b)$ is parabolic
$(|\mbox{trace}\ T_0(E_0,b)| =2)$ and hence $T_0(E_0,b)$ can be written as
$\left( \begin{array}{cc} 1 & d \\ 0 & 1 \\
  \end{array} \right)$ with $d\neq 0$. Because
\[\left\|\left( \begin{array}{cc} 1 & d \\ 0 & 1 \\ \end{array}
  \right)^{n_b}\right\|=
  \left\|\left( \begin{array}{cc} 1 & n_{b}d \\ 0 & 1 \\
  \end{array}\right)\right\|\leq
  C_d\ n_b,\] it follows that
\[\|\Phi_0(E_0,x,y)\|\leq C(E_0)n_b \leq C(E_0)|x-y|,\  \
\forall\ x,y\in\mathbb{N}.\] Therefore, by Corollary~\ref{DynamicCor2}
with $\alpha =1$, we obtain
\[\beta_{\delta_1^+}^-(0,q)\geq \frac{q-5}{2} ,\  \
\forall\ q>0.\]
\end{proof}

\

\subsection{The Thue-Morse Dirac model} This model is defined as
in~(\ref{DiracOperator}) by
\[\textbf{D}_{\omega}(m,c):=\textbf{D}_{0}(m,c)+ V_{\omega} \Id_2,\]
acting on
$\ell^2(\mathbb{N},\mathbb{C}^2)$ or
$L^2([0,\infty),\mathbb{C}^2)$, where $V_{\omega}$ is generated by the
Thue-Morse substitution on the
alphabet $\{a,b\}$ given by $S(a)=ab, S(b)=ba$. For more details
see~\cite{DLS,DST}. Let $\Omega_{\mathrm{TM}}$
be the associated subshift.

Since the boundedness of the transfer matrices in this case depends only
on the structure of the
potential and it is independent on the explicit form of these matrices, by
adapting a similar  model
\cite{DLS,DST} in the Schr\"odinger setting  we
obtain the following result (details omitted).

\

\begin{Lemma} There are $E_0\in\mathbb{R}$ and $C>0$ such that for every
$\omega\in\Omega_{\mathrm{TM}}$ and every $m\geq 0$,
\[\|\Phi_m^{\omega}(E_0,x,y)\|\leq C,\ \forall\ x,y\in\mathbb{N}
\ \ \mbox{or}\ \ \forall\ x,y\in [0,\infty).\]
\end{Lemma}

\

Thus, by Corollary~\ref{DynamicCor2} with $\alpha=0$, it follows that
\[\beta_{\omega,\psi}^-(m,q)\geq q-1,\] for every
$\omega\in\Omega_{\mathrm{TM}},\
q>0,\ m\geq 0$ and for every
$\psi=f\in\mathcal{H}_{E_0}$ in the continuous case
and $\psi=\delta_1^+$ in the
discrete case. This should be compared with Theorem~\ref{thmabqany}.

\

\subsection{The discrete Dirac model with Sturmian Potentials} We discuss
dynamical lower bounds for the
model
\[\textbf{D}_{\lambda,\omega,\theta}(m,c):=\textbf{D}_{0}(m,c)+
V_{{\lambda,\omega,\theta}} \Id_2 \] defined
by~(\ref{DiracOperator}) on
$\ell^2(\mathbb{N},\mathbb{C}^2)$, whose potential is given by
\[V_{\lambda,\omega,\theta}(x)=\lambda \chi_{[1-\omega,1)} (x\omega +\theta \
\ \mbox{mod} 1),\] where
$\lambda\neq 0$ is the coupling constant, $\omega\in (0,1)$ irrational is
the rotation number and
$\theta\in [0,1)$ is the phase. For more details on this potential in the
corresponding
Schr\"odinger case see~\cite{DL,IRT}.

Since the boundedness of the transfer matrices in this case depends only
on the structure of the
potential, again a direct adaptation of results in the Schr\"odinger
setting shows that

\

\begin{Lemma} Suppose $\omega$ is a number of bounded density. For every
$\lambda$, there are a constant $C>0$ and
$\alpha=\alpha(\lambda,\omega)>0$ such that for every
$\theta$ and every $E\in\sigma(\textbf{D}_{\lambda,\omega,\theta})$ we have
\[\|\Phi_{m,\lambda,\theta}^{\omega}(E,x,y)\|\leq C\ |x-y|^{\alpha},\] for
every $x,y\in\mathbb{N}$ and
any $m\geq 0$.
\end{Lemma}

\

Therefore, by  Corollary~\ref{DynamicCor1} with
$A=\sigma(\textbf{D}_{\lambda,\omega,\theta})$ (so $\mu_{+}^{m}(A)=1$),
it is found that for every
$\lambda, \theta$, the operator $\textbf{D}_{\lambda,\omega,\theta}$
satisfies
\[\beta_{\delta_1^+}^-(m,q)\geq \frac{q-3\alpha}{1+\alpha}\ ,\] for every
$q>0$ and any $m\geq 0$.

\

\section{Proof of Dynamical Bounds}
\label{ProofsSection} In this section the proof of
Theorem~\ref{DynamicTeor} will be presented. We first
gather some preliminary results that we will used in the proof.

For the operator $\textbf{D}(m,c),\ m\geq 0,$ on
$\ell^2(\mathbb{N},\mathbb{C}^2)$, we introduce the two-components Green's
function
\[\left(\begin{array}{c} G_{m}^+(z,n) \\  G_{m}^-(z,n)
\end{array}\right)= \left(\begin{array}{c}
\left\langle \delta_n^+,
\left(\textbf{D}(m,c)-z\right)^{-1}\delta_1^+ \right\rangle
\\  \left\langle \delta_n^-,
\left(\textbf{D}(m,c)-z\right)^{-1}\delta_1^+ \right\rangle
\end{array}\right), \ z\in \mathbb{C}\backslash\mathbb{R} , \] so that
\begin{equation} \label{Equ.AV} (\textbf{D}(m,c)-z)
\left(\begin{array}{c} G_{m}^+(z,n) \\  G_{m}^-(z,n)
\end{array}\right)= \delta_1^+(n) \ .
\end{equation} By using transfer matrices, one obtains for $n\geq 1$,
\begin{equation} \label{MatrGreen}
\left(\begin{array}{c} G_{m}^+(z,n) \\ G_{m}^-(z,n-1)
\end{array}\right)= \Phi_{m}(z,n,1)
\left(\begin{array}{c} G_{m}^+(z,1) \\  G_{m}^-(z,0)
\end{array}\right). \end{equation}

\

\begin{Lemma} \label{MomentsGreen} Let $\textbf{D}(m,c)$ be the
operator~(\ref{DiracOperator}). For
$z=E+i/T \ (T>0)$ and $m\geq 0$, one has
\[(i) \  \  A_{\delta_1^+}(m,T,q)=\frac{1}{\pi T}
            \sum_{n\in\mathbb{N}} n^q
            \int_{\mathbb{R}}\left(|G_{m}^+(z,n)|^2+|G_{m}^-(z,n)|^2\right)\
dE,\] in
the discrete case and
\[(ii) \  \  A_{f}(m,T,q)=\frac{1}{\pi T}
            \int_{0}^{\infty} x^q \int_{\mathbb{R}}
            \left\|\left(\textbf{D}(m,c)-z\right)^{-1}f(x)\right\|^2
            \ dE\ dx,\] for every $f\in L^2([0,\infty),\mathbb{C}^2)$, in
the continuous case.
\end{Lemma}

\begin{proof} The identity {\it(i)} follows by  Lemma 3.2 in~\cite{KKL}
adapted for the operator
$\textbf{D}(m,c)$ on $\ell^2(\mathbb{N},\mathbb{C}^2)$, and the identity
{\it(ii)} follows by  Lemma
2.3 in~\cite{DLS} applied to  $\textbf{D}(m,c)$ in
$L^2([0,\infty),\mathbb{C}^2)$.
\end{proof}

\

\begin{Lemma} \label{Form.Var.Energies} Let $E\in\mathbb{R}$, $N>0$,
$m\geq 0$ and consider
\[ L_m(N):=\sup_{0\leq x,y \leq N} \big
\| \Phi_{m}(E,x,y)\big \|\ .
\] Then, there is  $0<C_1<\infty$ such that for every
$\delta\in\mathbb{C}$ and $0\leq x,y \leq N$, one has
\[
\big \| \Phi_{m}(E+\delta,x,y)\big \|\leq L_{m}(N)\;
\exp\left[{\frac{|\delta|}{c}\left(\frac{|\delta|}{c}+C_1\right)L_m(N)|x-y|}\right].
\]
\end{Lemma}

\begin{proof} We consider the discrete case with $x,y\in\mathbb{N},\ x>y$
(the continuous case is
similar). An inductive argument shows that, for $\delta\in\mathbb{C}$ and
$m\geq 0$, we can write the identity
\[
\Phi_{m}(E+\delta,x,y)= \Phi_{m}(E,x,y)- \delta
\sum_{j=y}^{x-1} \Phi_{m}(E+\delta,x,j+1)\ B_{\delta}(E,j)\
\Phi_{m}(E,j,y)\ ,
\] with
\[B_{\delta}(E,j)= \frac{\delta}{c^2}
\left(\begin{array}{cc} 1 & 0 \\ 0 & 0
\end{array} \right) +\frac{1}{c} \left(
\begin{array}{cc} \frac{2}{c}(E-V(j)) & -1 \\ 1 & 0
\end{array} \right).\] By iteration, using the hypothesis and the above
identity, we obtain
\begin{eqnarray*}
\big \|\Phi_{m}(E+\delta,x,y)\big \| &\leq& L_{m}(N)
\left[1+\frac{|\delta|}{c}\left(\frac{|\delta|}{c}+C_1\right)
L_m(N)\right]^{x-y} \\ &\leq& L_{m}(N)\;
\exp\left[{\frac{|\delta|}{c}\left(\frac{|\delta|}{c}+C_1\right)L_m(N)(x-y)}\right],
\end{eqnarray*} for some $0<C_1<\infty$ and for $1\leq y < x \leq N$.
\end{proof}

\

The following result will be important for the proof of
Theorem~\ref{DynamicTeor} in the
continuous case; it is based on Lemmas~2.6 and 2.7 of \cite{DLS}.

\

\begin{Lemma} \label{Lema Continuo C.I.} Let $\textbf{D}(m,c)$ be the
operator defined
by~(\ref{DiracOperator}) on $L^2([0,\infty),\mathbb{C}^2)$. For
$z\in \mathbb{C}\backslash\mathbb{R}$, define
$u_{f,z}^{m}=\left(\textbf{D}(m,c)-z\right)^{-1}f$. Suppose
$E\in\mathbb{R}$ and
$0\neq f=\left(\begin{array}{c} f_+ \\ f_- \\ \end{array}\right)
   \in L^2([0,\infty),\mathbb{C}^2)$ with  $\mathrm{supp\,}f\subset [0,s]$
are such
that
\begin{equation} \label{Res.Conv.zero}
\liminf_{\delta \to 0^+} \left\{\|u_{f,z}^{m}(s)\|: z\in\mathbb{C}_+,\
|z-E|\leq \delta \right\}=0.
\end{equation} Then $f \notin \mathcal{H}_{E}$.
\end{Lemma}

\begin{proof} By~(\ref{Res.Conv.zero}) there exists a sequence
$(z_n)\subset\mathbb{C}_+$ with $z_n \to E$ and
$u_{f,z_n}^{m}(s)\rightarrow
\left(\begin{array}{c} 0 \\ 0 \end{array}\right)$ for $n \to \infty$.
Since $u_{f,z_n}^{m}(0)=
\left(\begin{array}{c} 0 \\ 0 \end{array}\right)$ for all $n$ and by
continuity, the inhomogeneous
equation
\begin{equation} \label{Eq.nAV}
\left(\textbf{D}(m,c)-E\right)
\left(\begin{array}{c} u_+ \\ u_- \end{array}\right)=
\left(\begin{array}{c} f_+ \\ f_- \end{array}\right)
\end{equation} has a solution
$v=\left(\begin{array}{c} v_+ \\ v_- \end{array}\right)$ with
$v(0)=v(s)=\left(\begin{array}{c} 0 \\ 0 \end{array}\right)$.

Let $Y(t)$ be the fundamental matrix of the homogeneous equation at
$x=~s$, i.e.,
\[
Y(t)=\left( \begin{array}{cc} v_+^N(t) & v_+^D(t) \\ \\
 v_-^N(t) & v_-^D(t) \\ \end{array} \right),
\]  where
$v^N=\left(\begin{array}{c} v_+^N \\ v_-^N \end{array}\right)$ and
$v^D=\left(\begin{array}{c} v_+^D \\ v_-^D \end{array}\right)$ are
solutions of the homogeneous equation
which satisfy
$v^N(s)=\left(\begin{array}{c} 1 \\ 0 \end{array}\right)$ and
$v^D(s)=\left(\begin{array}{c} 0 \\ 1 \end{array}\right)$. By writing
equation~(\ref{Eq.nAV}) as
\begin{eqnarray*}
\left(\begin{array}{c} u_+^\prime(x) \\ u_-^\prime(x) \end{array}\right)
\hspace{-0.2cm}&=&\hspace{-0.2cm}
\left( \begin{array}{cc} 0 & \frac{i}{c}(mc^2-V(x)+E) \\
 \frac{i}{c}(-mc^2-V(x)+E) & 0 \\ \end{array} \right)
 \left(\begin{array}{c} u_+(x) \\ u_-(x) \end{array}\right) \\
\hspace{-0.2cm}& &\hspace{-0.2cm} +\ \frac{i}{c}
 \left(\begin{array}{c} f_-(x) \\ f_+(x) \end{array}\right),
\end{eqnarray*} we have the variation of parameters formula
\[\left(\begin{array}{c} v_+(x) \\ v_-(x) \end{array}\right)= Y(x)
\int_{s}^{x} Y(t)^{-1}\ \frac{i}{c}
\left(\begin{array}{c} f_-(t) \\ f_+(t) \end{array}\right)\ dt.\]
Replacing $Y(t)$ in the above equation
 and considering $x=0$, we obtain
\begin{equation} \label{Eq.Produto}
0=v_+(0)=\frac{i}{c}\ [\overline{w_E},f] \  \  \  \mbox{and} \  \
\ 0=v_-(0)=\frac{i}{c}\ [\overline{v_E},f],
\end{equation}
where
\[w_E(t)=v_+^N(0)
 \left(\begin{array}{c} -v_+^D(t) \\ \\ v_-^D(t) \end{array}\right)
        +v_+^D(0)
 \left(\begin{array}{c} v_+^N(t) \\ \\ -v_-^N(t) \end{array}\right),\]
\[v_E(t)=v_-^N(0)
 \left(\begin{array}{c} -v_+^D(t) \\ \\ v_-^D(t) \end{array}\right)
        +v_-^D(0)
 \left(\begin{array}{c} v_+^N(t) \\ \\ -v_-^N(t) \end{array}\right)\] and
$f=\left(\begin{array}{c} f_+
\\ f_- \end{array}\right)$, with $f_+, f_-\neq 0$.

Now,
\[u^{1}(t):=-v_+^N(0)
 \left(\begin{array}{c} v_+^D(t) \\ \\ v_-^D(t) \end{array}\right)
        +v_+^D(0)
 \left(\begin{array}{c} v_+^N(t) \\ \\ v_-^N(t) \end{array}\right)\]
and
\[u^{2}(t):=-v_-^N(0)
 \left(\begin{array}{c} v_+^D(t) \\ \\ v_-^D(t) \end{array}\right)
        +v_-^D(0)
 \left(\begin{array}{c} v_+^N(t) \\ \\ v_-^N(t) \end{array}\right)\]
are solutions of equation
$\textbf{D}(m,c)
\left(\begin{array}{c} u_+ \\ u_- \end{array}\right)= E
\left(\begin{array}{c} u_+ \\ u_-
\end{array}\right)$ satisfying
$u^{1}(0)=\left(\begin{array}{c} 0 \\ -1 \end{array}\right)$ and
$u^{2}(0)=\left(\begin{array}{c} 1 \\ 0 \end{array}\right)$.
Thus, $u^{1}, u^{2}$ form a fundamental
system of solutions of
$\textbf{D}(m,c)u=Eu$ and it follows from (\ref{Eq.Produto}) that
\[\left[\overline{u^{i}},
\left(\begin{array}{c} f_+ \\ 0 \end{array}\right)\right] =
0=\left[\overline{u^{i}},
\left(\begin{array}{c} 0 \\ f_- \end{array}\right)\right] ,
\ i=1,2.\] Therefore, if
$f=\left(\begin{array}{c} f_+ \\ 0 \end{array}\right), f_+\neq 0,$ resp.\
$f=\left(\begin{array}{c} 0 \\ f_- \end{array}\right), f_-\neq 0,$ one has
\[\int_{0}^{s}u_+(t)f_+(t) dt =0,\ \mbox{resp.}\
\int_{0}^{s}u_-(t)f_-(t) dt =0,\] for every solution
$u=\left(\begin{array}{c} u_+ \\ u_- \end{array}\right)$ of
$\textbf{D}(m,c)u=Eu$. Hence, we conclude that $f \notin \mathcal{H}_{E}$.
\end{proof}

\

\begin{proof}{\bf (Theorem~\ref{DynamicTeor})}

\noindent $\bf (i)$ By Lemma~\ref{MomentsGreen}, we have for $T>0$,
\[A_{\delta_1^+}(m,T,q)=\frac{1}{\pi T}
     \sum_{n\in\mathbb{N}} n^q
     \int_{\mathbb{R}}\left(|G_{m}^+(E+i/T,n)|^2+
                       |G_{m}^-(E+i/T,n)|^2\right)\ dE.\] Define
$N(T):=T^{\frac{1}{1+\alpha}}$. By
hypothesis,
\[ L_m(N(T)):=\hspace{-0.2cm}\sup_{0\leq n,k \leq N(T)}
\big \| \Phi_{m}(E^{'},n,k)\big \| \leq C\ (N(T))^{\alpha},\ \
\forall\ E^{'}\in A(N(T)). \] By  Lemma~\ref{Form.Var.Energies}, we obtain
for every
$E\in B_2(T)$ and $1\leq n\leq N(T)$,
\[\big \| \Phi_{m}(E+i/T,n,1)\big \| \leq B\ (N(T))^{\alpha},\] with $B=C\
e^{\frac{3}{c}\left(\frac{3}{c}+C_1\right)C}$. For every $E\in B_2(T)$ and
$T$ sufficiently large, it
follows from~(\ref{MatrGreen}) and the  above estimate that
\begin{eqnarray} \label{Eq.8} & & \sum_{n\geq \frac{N(T)}{2}}
\left(|G_{m}^+(E+i/T,n)|^2+|G_{m}^-(E+i/T,n)|^2\right) \\ &\geq&
\sum_{n=\frac{N(T)}{2}+1}^{N(T)}
\left(|G_{m}^+(E+i/T,n)|^2+|G_{m}^-(E+i/T,n-1)|^2\right)\nonumber \\ &\geq&
\frac{B^{-2}}{4}(N(T))^{1-2\alpha}
(|G_{m}^+(E+i/T,2)|^2+|G_{m}^-(E+i/T,1)|^2 \nonumber \\ & &  +\
|G_{m}^+(E+i/T,1)|^2+|G_{m}^-(E+i/T,0)|^2). \nonumber
\end{eqnarray} Observe that
\[G_{m}^+(E+i/T,1)= \left\langle \delta_1^+,
\left(\textbf{D}(m,c)-E-i/T\right)^{-1}\delta_1^+\right\rangle =
F_m(E+i/T),\] where $F_m(z)$ is the
Borel transform of the spectral measure corresponding to the pair
$(\textbf{D}(m,c),\delta_1^+)$. Using  equation~(\ref{Equ.AV}) one shows that
\[|G_{m}^+(E+i/T,2)|^2+|G_{m}^-(E+i/T,1)|^2+|G_{m}^-(E+i/T,0)|^2
  \geq a > 0\] for some uniform constant $a$. Therefore, it follows
from~(\ref{Eq.8}) that for $T$
sufficiently large,
\begin{eqnarray*} & & \frac{1}{\pi T}\int_{\mathbb{R}}\sum_{n\geq
\frac{N(T)}{2}}
    \left(|G_{m}^+(E+i/T,n)|^2+|G_{m}^-(E+i/T,n)|^2\right)\ dE \\ &\geq&
\frac{\tilde{B}}{T}(N(T))^{1-2\alpha}\int_{B_2(T)}
       \left(1+ {\Im m}^2 F_m(E+i/T)\right)\ dE \\ &\geq&
\frac{\tilde{B}}{T}(N(T))^{1-2\alpha}\int_{B_2(T)}
       \left(\frac{1}{2}+ {\Im m} F_m(E+i/T)\right)\ dE ,
\end{eqnarray*} for some constant $\tilde{B}>0$. In the last step it was
used that
$1+ {\Im m}^2 F_m(z)\geq 2\ {\Im m} F_m(z)$.

For any set $S\subset\mathbb{R}$, denote by $S_{\epsilon}$ the
$\epsilon$-neighborhood of
$S$. It was shown
in~\cite{DST,KKL} that
\[\int_{S_{\epsilon}}{\Im\, m} F_m(E+i/T)\ dE \geq \frac{\pi}{2}\
  \mu_{+}^{m}(S).\] Thus, taking $S=B_1(T)$ we conclude that for
$T$ large enough,
\begin{eqnarray*} & & \hspace{-0.6cm} A_{\delta_1^+}(m,T,q)\ \geq \\
&\geq& \hspace{-0.2cm}
\frac{1}{\pi T}\left(\frac{N(T)}{2}\right)^{q}
\int_{\mathbb{R}}\sum_{n\geq \frac{N(T)}{2}}
\left(|G_{m}^+(E+i/T,n)|^2+|G_{m}^-(E+i/T,n)|^2\right)\ dE \\ &\geq&
\hspace{-0.2cm}
\frac{\tilde{C}}{T}(N(T))^{q+1-2\alpha}
\int_{B_2(T)} \left(1+ {\Im m} F_m(E+i/T)\right)\ dE \\ &\geq&
\hspace{-0.2cm}
\tilde{C}\ T^{\frac{q-3\alpha}{1+\alpha}}\
\left(|B_2(T)|+\mu_{+}^{m}(B_1(T))\right).
\end{eqnarray*}

\

\noindent $\bf (ii)$ As in Lemma~\ref{Lema Continuo C.I.} we write
$u_{f,z}^{m}=\left(\textbf{D}(m,c)-z\right)^{-1}f$. Let $s>0$ with
$\mathrm{supp\,}f\subset [0,s]$ and define
$N(T):=T^{\frac{1}{1+\alpha}}$. By Lemma~\ref{MomentsGreen}, we have for
$T>0$,
\begin{eqnarray} \label{Eq.10}
\hspace{0.8cm} A_{f}(m,T,q) \hspace{-0.2cm} &=& \hspace{-0.2cm}
    \frac{1}{\pi T}
    \int_{0}^{\infty} x^q \int_{\mathbb{R}}
    \|u_{f, E+i/T}^{m}(x)\|^2 \ dE\ dx \\ &\geq& \hspace{-0.2cm}
       \frac{1}{2\pi T} \sum_{n=s+1}^{\infty}(n-1)^q
       \int_{n-1}^{n+1} \int_{\mathbb{R}}
       \|u_{f, E+i/T}^{m}(x)\|^2 \ dE\ dx \nonumber \\ &\geq& \hspace{-0.2cm}
       \frac{1}{2\pi T} \sum_{n=s+1}^{\infty}(n-1)^q
       \int_{B_1(T)} \int_{n-1}^{n+1}
       \|u_{f, E+i/T}^{m}(x)\|^2 \ dx\ dE. \nonumber
\end{eqnarray}

Using the fact that $u_{f, E+i/T}^{m}$ is a solution of
$\textbf{D}(m,c)u=\left(E+\frac{i}{T}\right)u$ on $[n-1,n+1]$ and the
transfer matrices satisfy
$\|\Phi_m^{-1}\|=\|\Phi_m\|$, we obtain from~(\ref{Eq.10}) that
\begin{eqnarray*} & & \hspace{-0.8cm} A_{f}(m,T,q)\ \geq  \\
       &\displaystyle\frac{1}{2\pi T}& \hspace{-0.4cm}
       \sum_{n=s+1}^{\infty} \hspace{-0.1cm} (n-1)^q \hspace{-0.1cm}
       \int_{B_1(T)} \int_{n-1}^{n+1} \hspace{-0.1cm}
       \|\Phi_m(E+i/T,x,s)\|^{-2}
       \|u_{f, E+i/T}^{m}(s)\|^2 dx\ dE.
\end{eqnarray*} By hypothesis and Lemma~\ref{Form.Var.Energies}, it
follows that for $T$ large enough,
\begin{eqnarray*} A_{f}(m,T,q) \hspace{-0.3cm} &\geq& \hspace{-0.2cm}
       \frac{1}{\pi T} \hspace{-0.1cm}
       \sum_{n=\frac{N(T)}{2}+1}^{N(T)}\hspace{-0.1cm}
       \left(\frac{N(T)}{2}\right)^q
       \int_{B_1(T)}\hspace{-0.2cm} C_0 \, N(T)^{-2\alpha}
       \|u_{f, E+i/T}^{m}(s)\|^2\ dE \\
              &\geq& \hspace{-0.2cm}
       \frac{1}{\pi T} \left(\frac{N(T)}{2}\right)^{q+1}
       \hspace{-0.2cm}|B_1(T)|C_0 \,N(T)^{-2\alpha} \hspace{-0.2cm}
       \inf_{\mathrm{dist}(z,B_1(T))\leq\frac{1}{T}}\|u_{f, z}^{m}(s)\|^2,
\end{eqnarray*} for some constant $C_0>0$.

For every $f\in\mathcal{H}_{E_0}$ with $\mathrm{supp\,}f\subset [0,s]$,
Lemma~\ref{Lema Continuo C.I.} implies
that there exists $\kappa >0$ and $\delta >0$ satisfying
\[\inf \left\{\|u_{f,z}^{m}(s)\|^2 :
             z\in\mathbb{C}_+,\ |z-E_0|\leq \delta \right\}
       \geq \kappa .\] By hypothesis, $\mathrm{diam}(A(N))\longrightarrow
0$ as $N
\to\infty$ and
$E_0 \in A(N)$ for all $N$. Hence,
\[\inf \left\{\|u_{f,z}^{m}(s)\|^2 :
             \mathrm{dist}(z,B_1(T))\leq \frac{1}{T} \right\}
       \geq \kappa >0\] for $T$ sufficiently large.

Therefore, for $T$ large enough we obtain
\[A_{f}(m,T,q)\geq\frac{\tilde{C}}{T}\,N(T)^{q+1-2\alpha}\,|B_1(T)|
  =\tilde{C}\ T^{\frac{q-3\alpha}{1+\alpha}}\,|B_1(T)|. \]
The proof is complete.
\end{proof}

\

\


\begin{thebibliography}{99}

\bibitem{BD} Bjorken, S. D., Drell, J. D.: Relativistic quantum
mechanics. McGraw-Hill, New York (1965)

\bibitem{CadO} Carvalho, T. O., de Oliveira, C. R.: Critical energies
in random palindrome models. J. Math. Phys. {\bf 44}, 945--961 (2003)

\bibitem{DL} Damanik, D., Lenz, D.: Uniform spectral properties of
one-dimensional quasicrystals, II. The Lyapunov exponent. Lett. Math. Phys.
{\bf 50}, 245--257 (1999)

\bibitem{DLS} Damanik, D., Lenz, D., Stolz, G.: Lower transport bounds
for one-dimensional continuum Schr\"odinger operators. Math. Ann. {\bf 336},
361--389 (2006)

\bibitem{DST} Damanik, D., S\"{u}t\H{o}, A., Tcheremchantsev, S.:
Power-law bounds on transfer matrices and quantum dynamics in one
dimension II.
J. Funct. Anal. {\bf 216}, 362--387 (2004)

\bibitem{dOP1} de Oliveira, C. R., Prado, R. A.: Dynamical
delocalization for the 1D Bernoulli discrete Dirac operator.
J. Phys. A: Math. Gen. {\bf 38}, L115--L119 (2005)

\bibitem{dOP2} de Oliveira, C. R., Prado, R. A.: Spectral and
localization properties for the one-dimensional Bernoulli discrete Dirac
operator.
J. Math. Phys. {\bf 46}, 072105 17 pp (2005)

\bibitem{deOPqb} de Oliveira, C. R., Prado, R. A.: Quantum Hamiltonians
with quasi-ballistic dynamics and point spectrum.
J. Differential Equations {\bf 235}, 85--100 (2007)

\bibitem{GKT} Germinet, F., Kiselev, A., Tcheremchantsev, S.: Transfer
matrices and transport for 1D
Schr\"odinger operators. Ann. Inst. Fourier {\bf 54}, 787--830 (2004)

\bibitem{IRT} Iochum, B., Raymond, L., Testard, D.: Resistance of
one-dimensional quasicrystals.
Physica A {\bf 187}, 353--368 (1992)

\bibitem{JSBS} Jitomirskaya, S., Schulz-Baldes, H., Stolz, G.:
Delocalization in random polymer
models. Commun. Math. Phys. {\bf 233}, 27--48 (2003)

\bibitem{KKL} Killip, R., Kiselev, A., Last, Y.: Dynamical upper bounds
on wavepacket spreading. Am. J. Math. {\bf 125}, 1165--1198 (2003)

\bibitem{T} Thaller, B.: The Dirac equation. Springer-Verlag, Berlin (1991)


\end{thebibliography}
\end{document}